\documentstyle[graphicx,epsfig]{mn}

\newif\ifAMStwofonts

\newcommand{\texp}{t_{\mathrm{exp}}}

\ifoldfss
  \ifCUPmtlplainloaded \else
    \NewTextAlphabet{textbfit} {cmbxti10} {}
    \NewTextAlphabet{textbfss} {cmssbx10} {}
    \NewMathAlphabet{mathbfit} {cmbxti10} {} 
    \NewMathAlphabet{mathbfss} {cmssbx10} {} 
  \fi
  \ifAMStwofonts
    \ifCUPmtlplainloaded \else
      \NewSymbolFont{upmath} {eurm10}
      \NewSymbolFont{AMSa} {msam10}
      \NewMathSymbol{\upi}     {0}{upmath}{19}
      \NewMathSymbol{\umu}     {0}{upmath}{16}
      \NewMathSymbol{\upartial}{0}{upmath}{40}
      \NewMathSymbol{\leqslant}{3}{AMSa}{36}
      \NewMathSymbol{\geqslant}{3}{AMSa}{3E}

       \let\le=\leqslant
       
    \fi
  \fi
\fi 

\ifnfssone
  \newmathalphabet{\mathit}
  \addtoversion{normal}{\mathit}{cmr}{m}{it}
  \addtoversion{bold}{\mathit}{cmr}{bx}{it}
  \newmathalphabet{\mathbfit} 
  \addtoversion{normal}{\mathbfit}{cmr}{bx}{it}
  \addtoversion{bold}{\mathbfit}{cmr}{bx}{it}
  \newmathalphabet{\mathbfss} 
  \addtoversion{normal}{\mathbfss}{cmss}{bx}{n}
  \addtoversion{bold}{\mathbfss}{cmss}{bx}{n}
  \ifAMStwofonts
    \ifCUPmtlplainloaded \else
      \UseAMStwoboldmath
      \makeatletter
      \new@mathgroup\upmath@group
      \define@mathgroup\mv@normal\upmath@group{eur}{m}{n}
      \define@mathgroup\mv@bold\upmath@group{eur}{b}{n}
      \edef\UPM{\hexnumber\upmath@group}
      \new@mathgroup\amsa@group
      \define@mathgroup\mv@normal\amsa@group{msa}{m}{n}
      \define@mathgroup\mv@bold\amsa@group{msa}{m}{n}
      \edef\AMSa{\hexnumber\amsa@group}
      \makeatother
      \mathchardef\upi="0\UPM19
      \mathchardef\umu="0\UPM16
      \mathchardef\upartial="0\UPM40
      \mathchardef\leqslant="3\AMSa36
      \mathchardef\geqslant="3\AMSa3E

       \let\le=\leqslant

    \fi
  \fi
\fi 

\ifnfsstwo
  \DeclareMathAlphabet{\mathbfit}{OT1}{cmr}{bx}{it}
  \SetMathAlphabet\mathbfit{bold}{OT1}{cmr}{bx}{it}
  \DeclareMathAlphabet{\mathbfss}{OT1}{cmss}{bx}{n}
  \SetMathAlphabet\mathbfss{bold}{OT1}{cmss}{bx}{n}
  \ifAMStwofonts
    \ifCUPmtlplainloaded \else
      \DeclareSymbolFont{UPM}{U}{eur}{m}{n}
      \SetSymbolFont{UPM}{bold}{U}{eur}{b}{n}
      \DeclareSymbolFont{AMSa}{U}{msa}{m}{n}
      \DeclareMathSymbol{\upi}{0}{UPM}{"19}
      \DeclareMathSymbol{\umu}{0}{UPM}{"16}
      \DeclareMathSymbol{\upartial}{0}{UPM}{"40}
      \DeclareMathSymbol{\leqslant}{3}{AMSa}{"36}
      \DeclareMathSymbol{\geqslant}{3}{AMSa}{"3E}

       \let\le=\leqslant

    \fi
  \fi
\fi 

\ifCUPmtlplainloaded \else
  \ifAMStwofonts \else 
    \def\upi{\pi}
    \def\umu{\mu}
    \def\upartial{\partial}
  \fi
\fi

\title{The Effect of Gas Loss on the Formation of Bound Stellar Clusters}
\author[M. P. Geyer and A. Burkert]
       {M. P. Geyer and A. Burkert \\
        Max--Planck--Institute for Astronomy, K\"onigstuhl~17, 
		D-69117~Heidelberg, Germany}
\date{Accepted \dots
      Received \dots;
      in original form \dots}

\pagerange{\pageref{firstpage}--\pageref{lastpage}}
\pubyear{2000}

\begin{document}

\maketitle

\label{firstpage}

\begin{abstract}
The effect of gas ejection on the structure and binding energy of newly formed stellar clusters is investigated. The star formation efficiency (SFE), necessary for forming a gravitationally bound stellar cluster, is determined.

Two sets of numerical N--body simulations are presented: As a first simplified approach we treat the residual gas as an external potential. The gas expulsion is approximated by reducing the gas mass to zero on a given timescale, which is treated as a free parameter. 
In a second set of simulations we use smoothed particle hydrodynamics (SPH) to follow the dynamics of the outflowing residual gas self-consistently. 
We investigate cases where gas outflow is induced by an outwards propagating shock front and where the whole gas cloud is heated homogeneously, leading to ejection.

If the stars are in virial equilibrium with the gaseous environment initially,
bound clusters only form in regions where the local SFE is larger than 50\% or where the gas expulsion timescale is long compared to the dynamical timescale. A small initial velocity dispersion of the stars leads to a compaction of the cluster during the expulsion phase and reduces the SFE needed to form bound clusters to less than 10\%.
\end{abstract}

\begin{keywords}
globular clusters: general -- open clusters: general -- stars: formation.
\end{keywords}

 \section{Introduction}

During the formation of star clusters gas expulsion, caused by feedback of young massive stars, terminates the star formation epoch and can unbind the stellar system. Many mechanisms leading to gas loss exist, like ionizing radiation, stellar winds or supernova explosions. It is still uncertain which of those play the major role.

The gas expulsion reduces the binding energy of the cluster. Hills~\shortcite{Hills1980} showed, using analytic approximations, that a system of stars and gas loosing more than half of its mass in less than a crossing time will disrupt. Thus, to obtain bound stellar clusters, a star formation efficiency (SFE) $\epsilon = M_s / M_c > 0.5$ is needed, where $M_s$ and $M_c$ are the mass of the stellar component and of the initial gas cloud prior to star formation, respectively. As the typical SFEs are less than 10\%, the formation of gravitationally bound old open clusters and globular clusters is an interesting and yet unsolved problem.

Numerical simulations investigating the stability of young star clusters after gas expulsion have been done by Lada, Margulis \& Dearborn~\shortcite{Lada1984}. They showed that open star clusters, initially in virial equilibrium with the surrounding residual gas and containing up to 100~stars, can remain bound even if the SFE is as low as 30\%. In their simulations they treated the residual gas as a variable external potential added to that of the stars.
Goodwin~\shortcite{Goodwin1997} extended these simulations to globular clusters (GC), 
increasing the number of particles,
allowing for different gas expulsion mechanisms and including loss of stars due to the galactic tidal field.
Recently, Adams~\shortcite{Adams2000} presented a semi--analytic model for the formation of bound star clusters even for global SFEs much smaller than 50\%.

We present a new set of numerical simulations of the early evolution of young GCs, covering a broad range of SFEs and gas expulsion timescales. Because the typical collision dominated relaxation timescale of GCs is much larger than the dynamical timescale, we restrict ourselves to collisionless N--body calculations for the stellar component. We present two sets of simulations: 
As a first assumption, in Sect.~\ref{l_nbody} we treat the residual gas as an external potential. We investigate the dynamics of the cluster during and after gas expulsion and derive constraints for the SFE required to form a bound cluster. 
Using a combined N--body and hydrodynamic code, in Sect.~\ref{l_gas} we investigate the gas removal more self--consistently. Conclusions follow in Sect.~\ref{l_conclusion}.
\newpage

 \section[]{Gas Expulsion in Pure N--body\\* Simulations}

\label{l_nbody}
Our simulations start after the cluster has formed, but before the residual gas has been expelled.
The stars are represented by collisionless, equal mass N--body particles. The effect of the ejection of the residual gas on the stellar system is treated as a time variable external potential, similar to the approach of Lada et al.~\shortcite{Lada1984}. 

In the following, all quantities are given in dimensionless code units (gravitational constant $G=1$). Thus, taking typical globular cluster properties $\hat{M}=10^5\, \mathrm{M_{\odot}}$ and $\hat{R}=10\, \mathrm{pc}$ as mass and length units, respectively, we obtain a time unit $\hat{t}=(G\, \hat{\rho} )^{-1/2}=1.5\, 10^6\, \mathrm{yr}$ (equal to the dynamical or crossing timescale) with the density unit $\hat{\rho}=\hat{M}/\hat{R}^3$.

The N--body calculations are done using a hierarchical tree method~\cite{Barnes1986}.

 \subsection[]{Initial Configuration and the Gas Expulsion\\* Phase}

\label{l_nbody2}

To obtain a stable initial configuration, in a first step the stars are distributed according to a King~\shortcite{King1966} distribution function with total mass equal to that of the initial gas cloud. The potential is tabulated and is used for modelling the external gas potential during the simulation. Therefore, stars and gas have equal density distributions. Finally, the mass of the stars and the gas are scaled according to the SFE.
Now the stars are in virial equilibrium within the sum of their own potential and the potential of the gaseous component.
The parameters of the different models are given in Table~\ref{tab1}. For each model, the SFE is varied between $0.15$ and $0.80$.
\begin{table}
	\caption{Parameters of the initial configurations (N--body)}
	\label{tab1}
	\begin{center}\begin{tabular}{llllll}
	 model	&	$W_0$ & $t_c$ & $R_t$ & $N$ & $\delta$\\ \hline
	 N1 & $3.0$ & 0.28 & 1.26 & $1000$ & $0.1$\\  
	 N2 & $5.1$ & 0.17 & 1.33 & $1000$ & $0.1$\\  
	 N3 & $3.0$ & 0.28 & 1.26 & $4000$ & $0.05$\\ \hline 
	\end{tabular}\end{center}
	$W_0=\psi(0)/\sigma^2$: scaled central potential of the King profile,
	see Binney \& Tremaine~\shortcite{Binney1987}; 
	$t_c$: crossing time at half--mass radius;
	$R_t$: tidal radius of King profile; 
	$N$: number of particles used in simulation; 
	$\delta$: numerical (Plummer) smoothing length;
	total mass (stars and gas) of all models was set to 1;
	all quantities are given in dimensionless
	code units (see text)
\end{table}
	
To test whether the initial system is in virial equilibrium, several calculations without gas expulsion are performed. The density distributions are well conserved.

We simulate the gas expulsion starting at time $t=t_0$ by multiplying the external potential by a time dependent factor 
\begin{equation}
	\xi = \left\{
	  \begin{array}{lcl}
	  	1 & & t<t_0\\
		1-(t-t_0)/\texp & \mathrm{if} & t_0<t<t_0+\texp \\
		0 & & t>t_0+\texp
	   \end{array} \right. ,
\end{equation}
where $\texp$ is the time that is needed to drive the gas out of the system (gas expulsion time).

We can estimate the order of the gas expulsion time: The isothermal sound speed of a molecular cloud gas with temperature $T=10\, \mathrm{K}$ and molecular weight $\mu=2.36$
is $a=\sqrt{R_{\mathrm{gas}}\, T / \mu}=0.19\, \mathrm{km}\; \mathrm{s}^{-1}$, where $R_{\mathrm{gas}}$ is the gas constant.
If we consider a disruptive process that starts at the centre of a cloud as given by model~N1 and travels outwards with sound speed, it will need approximately a time of $\texp  = R_t\, a^{-1}\approx 6.6\, 10^6\, \mathrm{yr}$ (or $\texp \approx 4$, given the dimensionless code units above) to reach the edge of the cloud. Fast processes (e.g. supernova explosions) may remove the gas on shorter timescales.
The gas expulsion time will therefore presumably be of the order of a dynamical time which is equal to the unit of time. In the simulations we use $\texp =0, 2, 4$ and $10$, which are equal to $0, 7, 14$ and $36$ crossing times at half-mass radius of model~N1 and~N3
and $0, 12, 24$ and $59$ crossing times of model~N2.

 \subsection[]{Dynamics of the Cluster During and After\\* Gas Expulsion}

\begin{figure}
	\includegraphics[width=84mm,height=84mm]{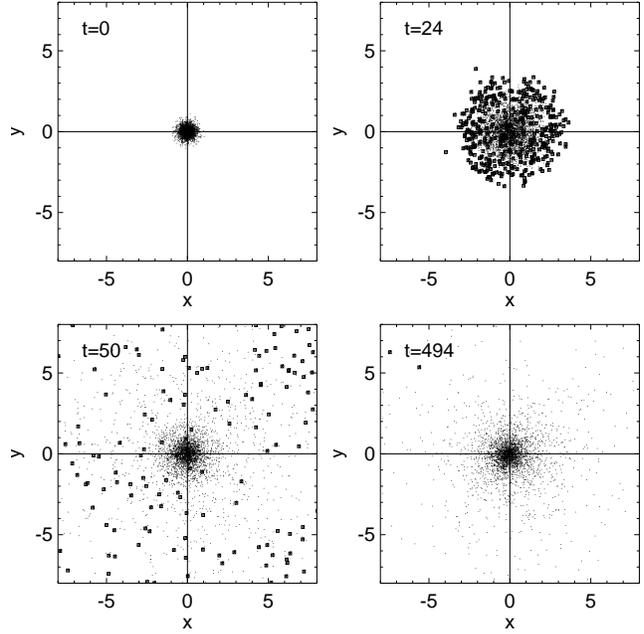}
	\caption{Time evolution of model~N3. 
	     The time $t$ is given in dimensionless units. 
	     The SFE is $\epsilon =0.4$,
		 expulsion time $\texp =2$. The plots show the N--body
		 particles (small dots) projected onto the x--y--plane; 
		 unbound particles are indicated by squares.
			 }
	\label{abb1}
\end{figure}

The typical evolution of the N--body part of a cluster with $\texp=2$ ($7$ crossing timescales) is displayed in Fig.~\ref{abb1}.
Starting at $t=20$ the external potential is slowly reduced to zero as described in the previous section. The cluster expands. A certain amount of stars gets unbound and starts leaving the system. The bound ones relax after the gas has been completely removed, forming a broader configuration.
A particle is believed to be unbound if its total energy (kinetic energy plus potential energy) is positive. This criterion is different from Goodwin~\shortcite{Goodwin1997}, who marked all stars outside a given tidal radius as unbound.

\begin{figure}
	\includegraphics[width=84mm,height=84mm]{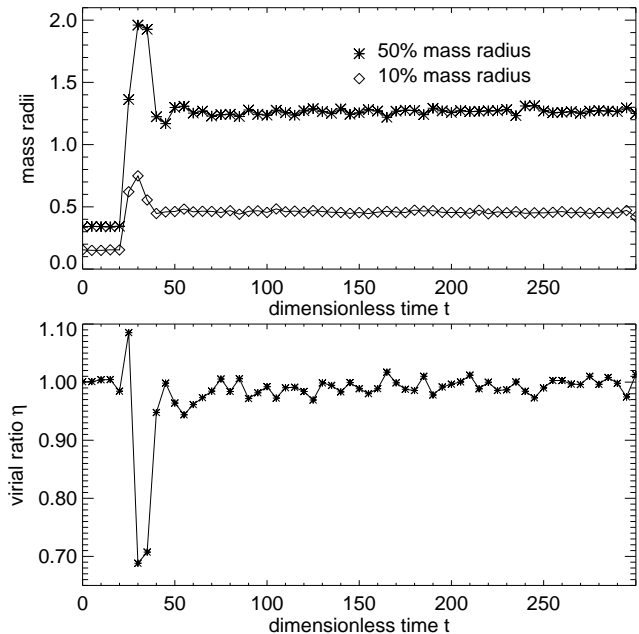}
	\caption{Time evolution of the mass radii (upper panel) and the 
	         virial ratio $\eta$ (lower panel)
         	 of the system shown in Fig.~\ref{abb1}
			 }
	\label{abb2}
\end{figure}

Fig.~\ref{abb2} shows the evolution of the Lagrangian radii containing 10\% and 50\% of the {\it current bound\/} mass of the system and the virial ratio $\eta=-2\, E_{\mathrm{kin}}/E_{\mathrm{pot}}$ of the bound particles. 
The constant mass radii and virial ratios before gas expulsion show that initially the system is indeed in virial equilibrium.
When the superimposed gas potential decreases ($t>20$), the mass radii increase rapidly and relax for $t>\texp $. This behaviour is also reflected in the virial ratio. The system achieves a more extended equilibrium state.

The radial expansion factor of the cluster can be estimated as follows:
In the adiabatic case (expulsion timescale is much longer than the crossing time) $R\, \cdot M$ is constant and therefore~\cite{Hills1980,Mathieu1983} the ratio of final to initial radius is
\begin{equation}
	\frac{R_f}{R_i}=\frac{M_c}{M_s}=\frac{1}{\epsilon}\quad \mathrm{with}\quad
	0 < \epsilon \le 1.
	\label{m1}
\end{equation}
On the other hand, if the expulsion time is short compared to the crossing time,
we can apply conservation of kinetic energy per particle during the ejection of the gas.
Hills~\shortcite{Hills1980} and Mathieu~\shortcite{Mathieu1983} obtained
\begin{equation}
	\frac{R_f}{R_i}=\frac{\epsilon}{2\, \epsilon -1}\quad \mathrm{with}\quad
	\frac{1}{2} < \epsilon \le 1.
	\label{m2}
\end{equation}
If $\epsilon \le 0.5$ the final system is unbound.

\begin{figure}
	\includegraphics[width=84mm,height=54.6mm]{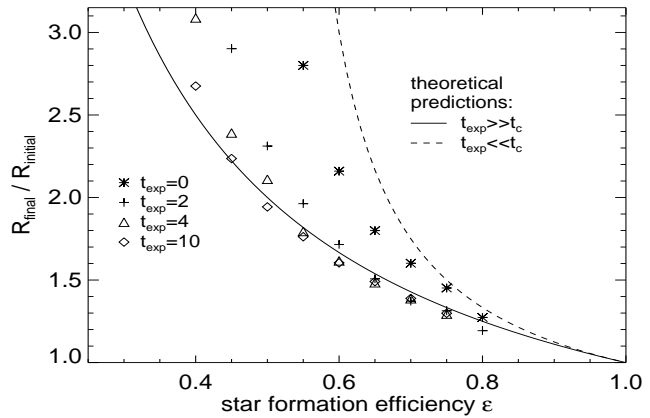}
	\caption{Ratio of the final to the initial half--mass radii versus SFE of
	         the simulations (model N3) for various 
			 expulsion timescales $\texp $
			 }
	\label{abb3}
\end{figure}

Fig.~\ref{abb3} shows the ratio of the final to the initial half--mass radii of the {\it bound} particles at the end of the simulations, compared to the theoretical predictions of Eq.~\ref{m1} and~\ref{m2}. 
If stars are lost and the bound mass of the cluster is not conserved, Eq.~\ref{m1} and~\ref{m2} are not strictly valid any more and discrepancies to the analytic approximations occur.

As expected, the simulations with long gas expulsion timescales fit well the solid curve, representing the adiabatic case.
The faster the gas expulsion, the larger is the ratio of the final to the initial radius compared to the theoretical result.

The models with fast gas expulsion follow the dashed curve well for high SFEs. For low SFEs, the ratio of final to initial radii is smaller than the analytical prediction, which
emphasizes that the divergence for $\epsilon=0.5$ does not occur
in numeric simulations: The final radius is
decreased by excluding the unbound particles which are located preferentially at high radii. Additional, the outgoing particles reduce the total energy of the remaining system and leave behind a tighter bound core.

 \subsection{Constraints on the Star Formation Efficiency}

After the gas expulsion the system of the remaining bodies relaxes
again (Fig.~\ref{abb2}). 

\begin{figure}
	\includegraphics[width=84mm,height=100.8mm]{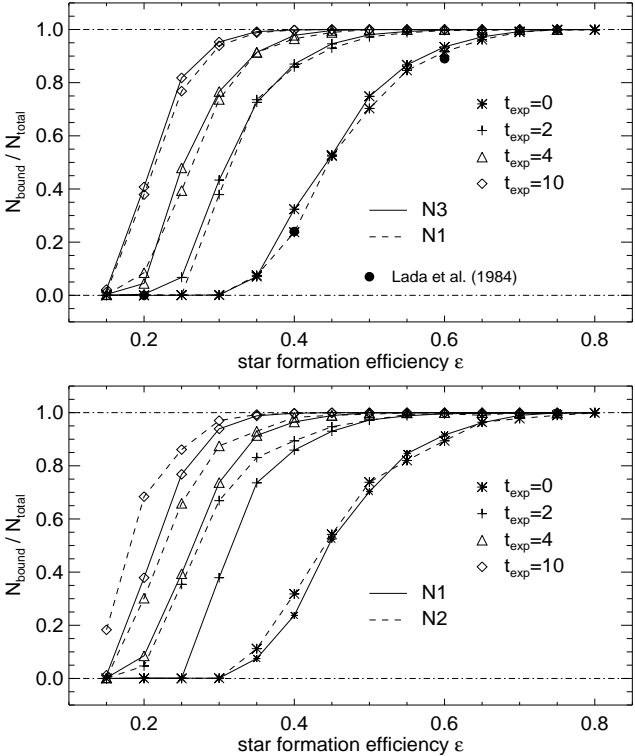}
	\caption{Ratio of the number of bound stars to the initial number of stars in              the relaxed system after gas expulsion.
	         Upper panel: models N1 and N3; 
			 lower panel: models N1 and N2; each symbol represents one run
			 with given SFE and gas expulsion time; large dots are results taken
			 from Lada et al.~\shortcite{Lada1984}, Fig.~2 therein
			 }
	\label{abb4}
\end{figure}

Fig.~\ref{abb4} shows the ratio of the number of finally bound stars
to the initial number of stars for various SFEs and gas expulsion timescales.
The upper panel provides a resolution study of the runs N1 and N3 with 1000 and 4000 particles, respectively. Within the uncertainties they are indistinguishable. 
Large dots show results from Lada et al.~\shortcite{Lada1984} obtained from simulations with 50~(!) stars. As can be seen, the number of particles used does not influence the results.

The lower panel compares the initially more concentrated King model (N2)
to the less concentrated one (N1). We find that
the curves of model N2 with $\texp>0$ are shifted to lower star formation efficiencies or
higher ratios of bound stars, respectively.
This is due to the lower half--mass crossing time of the more
concentrated cluster N2 (Table~\ref{tab1}), confirming that only the ratio of the expulsion timescale to the crossing time is important. Thus, more concentrated clusters have a larger chance to survive.
In the case of instantaneous gas expulsion ($\texp=0$) the models N1 and N2
yield the same curve.

The ratio of bound to unbound stars gives a 
threshold for the SFE $\epsilon$ necessary to 
form bound clusters. 
In case of instantaneous gas expulsion (see Fig.~\ref{abb4}) the curves are centered around $\epsilon=0.45$, which is somewhat less than the theoretical
limit $\epsilon=0.5$ for bound clusters given by Hills~\shortcite{Hills1980}.

Adams~\shortcite{Adams2000} recently gave analytic approximations for the
dependency of the number of bound stars on the SFE in the case of
instantaneous gas expulsion. For a SFE $\epsilon=0.5$
about 73\% of the stars are kept, in good agreement to our results from
Fig.~\ref{abb4} ($\texp=0$). However, our results show
a stronger dependence of mass loss on the SFE $\epsilon$.
Contrary to Adams~\shortcite{Adams2000},
star clusters with a SFE lower than $\epsilon=0.4$ are dissolved in our
simulations. This discrepancy can be understood from the fact that Adams uses density distributions of gas and stars with very different concentrations. Thus, even if the global SFE is small, the local SFE in the region of star formation could be as high as 90\%, leading to a bound system.

The number of finally bound stars increases with the gas expulsion
time. In order for more than 50\% of all the stars to remain bound the SFE must be equal to 45\%, 30\%, 25\% and 20\% for gas expulsion times $\texp =$0, 2, 4 and 10, respectively.
The Galactic average SFE in giant molecular clouds is of the order of a few percent~\cite{Myers1986,Williams1997}. Koo~\shortcite{Koo1999} has observed SFEs up to 15\% in the star forming--region W51B, maybe due to shock--interaction with a spiral density wave.
Given these SFEs, unrealistic high gas expulsion timescales are required to obtain bound clusters. Only few clouds show SFEs up to 30\% or 40\%~(Lada 1992) and may stay bound.

 \section[]{Combined N--body and\\* Hydrodynamical Simulations}

\label{l_gas}
In Sect.~\ref{l_nbody}, the effect of gas removal is treated as a time variable external potential.
To describe the physics more properly, we extend our simulations using smoothed particle hydrodynamics (SPH, see Monaghan~1992 for a review). We are using an SPH--code with variable smoothing length, individual particle timesteps (Bate, Bonnell \& Price~1995) and collisionless N--body particles, which was made available by Matthew Bate.

The conversion from code units to physical units is the same as in Sect.~\ref{l_nbody2}. Additionally, we describe the internal energy of the gas with a dimensionless temperature that scales with $\hat{T}= G\, \hat{M}\, \mu\, \hat{R}^{-1}\, R_{\mathrm{gas}}^{-1}\, \gamma^{-1}$, where $\mu$ is the mean molecular weight, $R_{\mathrm{gas}}$ is the gas constant and $\gamma$ the constant adiabatic exponent. For an ideal gas with $\gamma=3/2$ and $\mu=1.0$, we have $\hat{T}=3.5\,10^3\, \mathrm{K}$.

 \subsection[]{Initial Configuration and Models for Gas\\* Expulsion}

Murray \& Lin~\shortcite{Murray1992} proposed as initial conditions pressure--confined protocluster clouds. As a reasonable initial configuration we therefore adopt a pressure--confined, isothermal Bonnor--Ebert (BE) sphere~\cite{Ebert1955,Bonnor1956}, see Table~\ref{tab2}.
\begin{table}
	\caption{Parameters of the initial configurations including gas
	dynamics (N--body \& SPH)}
	\label{tab2}
	\begin{center}\begin{tabular}{llllllll}
	 model	& $r_h$	 & $t_c$ &  $N$ & $T$ & $\delta$ & $R$ &$\xi_b$\\ \hline
	 S1 &  $0.7$ & $0.8$ & $2\times 4000$ & $0.76$ & $0.01$ & $1.0$ & $4.0$\\  
	 S2 &  $0.7$ & $0.8$ & $2\times 11045$ & $0.76$ & $0.05$ & $1.0$ & $4.0$\\  
	\end{tabular}\end{center}
	$r_h$: half--mass radius;
	$t_c$: crossing time at half--mass radius;
	$N$: number of particles;	
	$T$: gas temperature; $\delta$: numerical (Plummer) smoothing length  
	(N--body part only); $R$: cut--off radius;
	$\xi_b$: dimensionless cut-off radius, see Bonnor~\shortcite{Bonnor1956};
	total mass (stars and gas) of all models was set to 1;
	all quantities are given in dimensionless
	code units (see text)
\end{table}

To get a combined system of gas and stars, we add N--body particles with an equal density distribution, scaling the masses of the stars and the gas according to the SFE $\epsilon$. The velocity dispersion of the particles is chosen according to the temperature of the gas. 

Isothermal spheres extend to infinity. In order to get a finite configuration, the gas density is set to zero at an arbitrary radius. To stabilize the gas sphere an external pressure is applied. This is not possible for particle systems. Therefore the velocity dispersion of the N--body particles is decreased and the system is allowed to relax until a stable configuration is obtained. 

\begin{figure}
	\includegraphics[width=84mm,height=42mm]{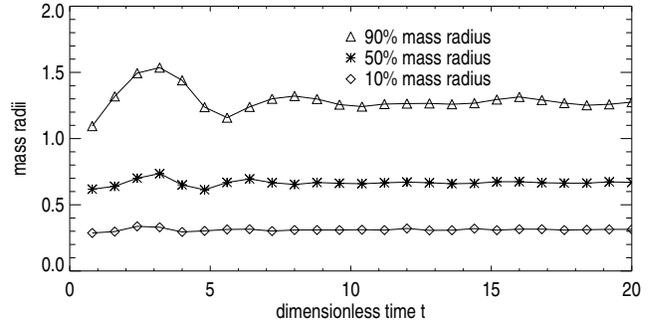}
	\caption{Evolution of the mass radii of a combined N--body and
	         SPH simulation. At the end of the stability test the 
			 system is in equilibrium
			 }
	\label{abb5}
\end{figure}

Fig.~\ref{abb5} shows the evolution of one typical setup prior to gas removal. About 2\% of the N--body particles are lost, but after some oscillations the main part achieves an equilibrium state. This configuration is then used as initial model for following investigations.
The different models used are shown in Table~\ref{tab2}.

As the processes of gas expulsion are not well understood, we choose two simplified scenarios: In our first model we heat up the whole gas cloud. As a result, the gas starts to expand and is removed. Such a situation might be caused by stellar winds or ionizing radiation.
In our second scenario we heat up a small inner core, creating an outward propagating shock front disrupting the gas cloud. 
Such shock fronts may be generated by combined supernova explosions and winds from central high--mass stars. The formation of supershells and their ability to disrupt the cloud are discussed in various papers with regard to chemical self--enrichment of GCs~\cite{Morgan1989,Brown1995,Parmentier1999}. Goodwin, Pearce \& Thomas~\shortcite{Goodwin2000} investigate single supernovae in gas clouds.
In our simulations the heating of the cloud, determining the gas expulsion timescale, must be sufficient to expel all the gas.

 \subsection[]{Evolution of the Cluster During and After\\* Gas Expulsion}

\begin{figure}
	\includegraphics[width=84mm,height=84mm]{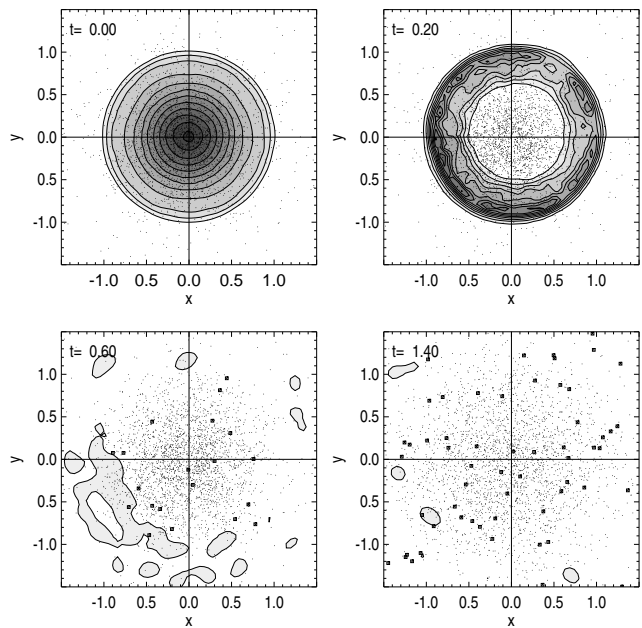}
	\caption{Time evolution of model S2 during the gas expulsion phase
	     (dimensionless time $t$).
	     The SFE is
	     $\epsilon =0.4$ and central heating is applied. The plots show the
		 N--body
		 particles (small dots) projected onto the
		 x--y--plane (only every 4th particle is shown); 
		 unbound particles are indicated by squares.		 
			 The contour lines indicate the gas density in the
			 x--y--plane in steps of 0.025.
			 }
	\label{abb6}
\end{figure}

Fig.~\ref{abb6} shows the time evolution of a system with gas ejection in a supershell. Once the internal pressure of the expanding gas is smaller than the adopted external pressure, the outward propagating shell becomes unstable and develops substructures. At that stage, less than two crossing times after the gas removal started, the gas density in the cluster region is so low that its gravitational effect on the stellar component is negligible. We therefore remove the gas and follow the evolution of the stars alone.
The globally heated cloud models show a very similar evolution and are treated in the same manner.

\begin{figure}
	\includegraphics[width=84mm,height=100.8mm]{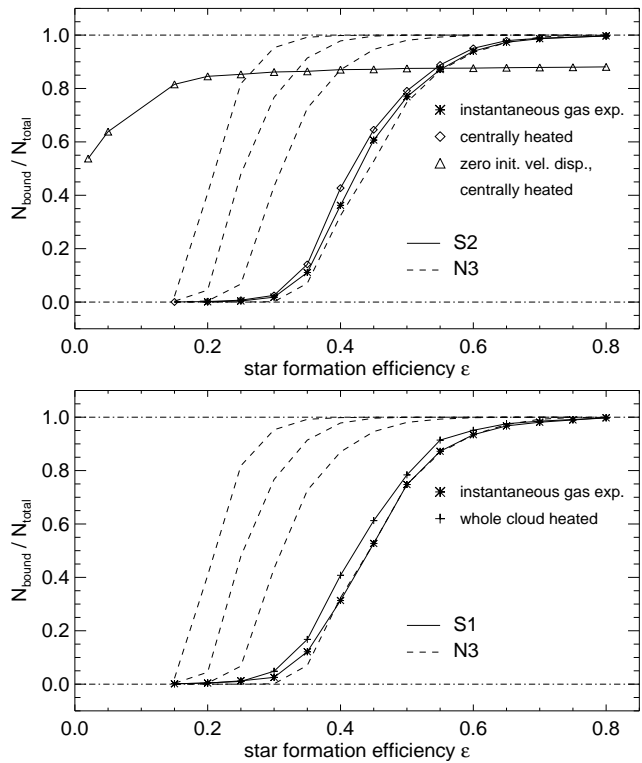}
	\caption{Ratio of the number of bound stars to the initial number of stars in              the relaxed system after gas expulsion;
			 dashed lines
			 indicate the results from Fig.~\ref{abb4}, model N3			 }
	\label{abb7}
\end{figure}

Fig.~\ref{abb7} corresponds to Fig.~\ref{abb4} in the pure N--body case. For comparison, the dashed curves reproduce the results of the pure N--body simulations.

For instantaneous gas expulsion (all the gas particles are removed at once), corresponding to the case $\texp =0$ in Sect.~\ref{l_nbody}, the simulations using the BE sphere as initial configuration are in very good agreement with the simulations using a King density distribution. The slightly higher crossing times of the BE sphere (see Table~\ref{tab2}) may cause the small differences for small SFEs: the simulation was not run long enough for all particles to get unbound. We therefore conclude that the density distribution has no influence on the number of bound particles after gas expulsion, at least if $\texp =0$.

The crosses in Fig.~\ref{abb7}, lower panel, show system S1 where the temperature of the whole cloud was increased by a factor of 10 with respect to the equilibrium model. The number of bound stars increases slightly compared to the case with instantaneous gas expulsion. The diamonds in the upper panel show the results for a cloud centrally heated to $T=152$ (system S2). Compared to the first model S1, no significant differences are visible. Again, the number of bound stars increases slightly.
However, the gas expulsion process in both cases is much faster than the timescales adopted in Sect.~\ref{l_nbody}. Therefore, also in more realistic cases, high SFEs are needed to sustain a bound star cluster.

One way out may be a collapse of the star cluster before the gas is completely expelled, leading to a higher ``effective SFE''. Lada et al.~\shortcite{Lada1984} and Verschueren~\shortcite{Verschueren1990} proposed a low or zero initial velocity dispersion to explain the collaps. Saiyadpour, Deiss \& Kegel~\shortcite{Saiyadpour1997} considered the effect of dynamical friction on the stellar cluster.

We implement the first approach by setting the initial velocity dispersion of
the stars to zero and adopting gas expulsion by heating the whole gas cloud. The cluster virializes within a small radius and
only few stars that gain velocities higher than the escape velocity of the cluster are ejected.
For a wide SFE range the percentage of bound stars at the end of the
simulations is nearly constant and
higher than $80\%$ (Fig.~\ref{abb7}, upper panel, triangles). 
Even our run with the lowest SFE $\epsilon=0.02$ leads to a remaining
bound system.
We conclude that this scenario can easily explain the
formation of bound clusters. 
However, if the gas expulsion is delayed one dynamical time scale or longer,
the stars virialize in a smaller volume. We then basically get back to the
case of an initially virialized system with a higher local
SFE than in the beginning. The system may break up again and the number of bound stars will be less than in the scenario with low velocity dispersion, but higher than in the initially virialized case.

 \section{Conclusions \& Outlook}

\label{l_conclusion}

N--body and combined N--body \& SPH calculations to investigate the influence 
of the residual gas expulsion on the stellar part in 
star forming regions are presented.

We show that in the case of instantaneous gas expulsion,
clusters with SFEs greater than $\epsilon=0.45$ can keep more than 50\%
of the initial stars. Clusters with SFEs less than $\epsilon=0.40$ are dissolved. Different concentrations of the initial models show no effect at all on the number of bound stars if the gas is expelled instantaneous.

We confirm that gas expulsion timescales which are several times longer than the crossing time of the GC can decrease the SFE needed to sustain bound clusters considerably.

However, our simulations including the proper dynamics of the residual gas show
that in order to destroy the whole cloud by global heating
or by supershells the gas expulsion must take place on a short timescale, requiring a high SFE. Only few star forming regions show such high SFEs.
 
We demonstrate that models with stars having an initial zero velocity dispersion
lead to a compaction of the cluster and can explain bound systems even in low SFE regions: For SFEs as low as $\epsilon=0.15$ more than 80\% of the stars
stay bound. Bound systems are obtained even with SFEs lower than 10\%.
For future investigations it is essential to know the velocity dispersion of newly born stars in clusters.

\bigskip
{\it Acknowledgments.\/} We would like to thank Matthew Bate for making available his combined N--body \& SPH code. Also thanks to Pavel Kroupa for
his very useful comments.


\bsp

\label{lastpage}

\end{document}